# Long-Term Orbit Dynamics of Decommissioned Geostationary Satellites

Simone Proietti[a], Roberto Flores[b,c], Elena Fantino[b], and Mauro Pontani[d*]

[a] Faculty of Civil and Industrial Engineering, Sapienza University of Rome, via Eudossiana 18, 00186 Rome, Italy

[b] Department of Aerospace Engineering, Khalifa University of Science and Technology, Abu Dhabi, United Arab Emirates, P.O. Box 127788

[c] Centre Internacional de Mètodes Numèrics en Enginyeria (CIMNE), Gran Capità s/n, 08034 Barcelona, Spain

[d] Department of Astronautical, Electrical, and Energy Engineering, Sapienza University of Rome, via Salaria 851, 00138 Rome, Italy

[*] Corresponding author

Email simone.proietti14@gmail.com, rflores@cimne.upc.edu, elena.fantino@ku.ac.ae, mauro.pontani@uniroma1.it

**Abstract**

In nominal mission scenarios, geostationary satellites perform end-of-life orbit maneuvers to reach suitable disposal orbits, where they do not interfere with operational satellites. This research investigates the long-term orbit evolution of decommissioned geostationary satellite under the assumption that the disposal maneuver does not occur and the orbit evolves with no control. The dynamical model accounts for all the relevant harmonics of the terrestrial gravity field at the typical altitude of geostationary orbits, as well as solar radiation pressure and third-body perturbations caused by the Moon and the Sun. Orbit propagations are performed using two algorithms based on different equations of motion and numerical integration methods: (i) Gauss planetary equations for modified equinoctial elements with a Runge-Kutta numerical integration scheme based on 8-7$^{th}$-order Dorman and Prince formulas; (ii) Cartesian state equations of motion in an Earth-fixed frame with a Runge-Kutta Fehlberg 7/8 integration scheme. The numerical results exhibit excellent agreement over integration times of decades. Some well-known phenomena emerge, such as the longitudinal drift due to the resonance between the orbital motion and Earth's rotation, attributable to the $J_{22}$ term of the geopotential. In addition, the third-body perturbation due to Sun and Moon causes two major effects: (a) a precession of the orbital plane, and (b) complex longitudinal dynamics. This study proposes an analytical approach for the prediction of the precessional motion and show its agreement with the (more accurate) orbit evolution obtained numerically. Moreover, long-term orbit propagations show that the above mentioned complex longitudinal dynamics persists over time scales of several decades. Frequent and unpredictable migrations toward different longitude regions occur, in contrast with the known effects due only to the perturbative action of $J_{22}$.

## 1. Introduction

In the last decades, the space debris population has grown rapidly [1], creating a serious hazard for existing spacecraft and future space missions. As a result, multiple studies and proposals have appeared in the scientific literature that address the problem of space debris mitigation and removal [2-6]. Moreover, the International Academy Debris Committee has supplied several recommendations to space ventures and agencies to avoid the increasing deterioration of the space environment [7]. Specifically, two regions are critically crowded: (a) the spherical shell corresponding to low Earth orbits, particularly at altitudes between 400 and 800 km, and (b) the circular ring around the geostationary (GS) orbit [8]. The latter typically hosts telecommunications and remote sensing satellites. Nowadays, these spacecraft are equipped with the propellant needed to reach adequate graveyard orbits at the end of their operational life and thus avoid interfering with active satellites. However, in the sixties, seventies and early eighties GS spacecraft used to be abandoned in the synchronous orbit at the end of operations, which led to the accumulation of an enormous amount of debris in the region. Even today, if the propulsion system fails, the satellite drifts uncontrolled under the effects of orbital perturbations. Therefore, a thorough understanding of the long-term orbital dynamics and the ensuing hazard for current and future operations in the GS region is of paramount importance.

For GS satellites, the $J_{22}$ term, related to the ellipticity of the terrestrial equator, has a dominant effect due to resonance between the orbital motion and Earth's rotation. This gives rise to four equilibrium positions, two stable and two unstable, at specific geographical longitudes. The perturbed motion due only to $J_{22}$ is completely predictable, and consists of two qualitatively different dynamical behaviors: (i) nonlinear longitudinal oscillations about the stable equilibrium positions or (ii) longitudinal circulation (either westward or eastward). Separation curves, termed separatrices, divide behaviors (i) and (ii).

Early research, due to Vashkov'yak and Lidov [9] pointed out the existence of librational motion encircling both the stable points. This effect was attributed to the action of higher tesseral harmonics of the geopotential, which cause a substantial alteration of the separatrices. Moreover, the same authors provided estimations of the libration periods in the regions that surround stable points. In 1985 Hechler [10] addressed the issue of unregulated use of the GS orbit and analyzed the hazard due to abandoned objects and the probability of collisions with operational satellites. Orbit propagations based on a first-order perturbation theory showed uncontrolled objects passing through the GS region. Simulations spanning a few decades revealed a precession of the pole of the orbit with a period of about 53 years around a point about 7.3 degrees from the Earth's spin axis in the direction of the pole of the ecliptic, which the author attributed to the joint effects of Earth's oblateness and lunisolar attraction. Hechler [10], Van der Ha [11], and Allan and Cooke [12,13] had already analyzed this effect using the leading terms of the lunisolar perturbing function in an analytical averaging over the mean anomaly for both the satellite and the perturbing bodies.

Other investigations focused on the orbital evolution of near-GS satellites, particularly objects disposed at 100 km above the GS orbit or higher. Van der Ha [14] developed an averaged-orbit model to describe the evolution of near-GS orbits over time intervals of a few years and validated it through the observations of the GEOS-2 satellite, which at the time was orbiting 260 km above the GS orbit. The objective was to have a tool for quick orbit predictions of the orbital evolution of disposed satellites. The model accounted for the secular contributions of the lunisolar perturbations and the gravitational harmonics of the Earth up to the fourth degree, while the solar radiation forces were determined assuming constant material parameters. Friesen [15] illustrated the outcome of numerical propagations of GS and near-GS satellites over intervals of 10 and 1000 years, respectively, to check for possible future intersections with the synchronous ring. The physical model included a fourth-degree gravity field and solar radiation pressure, and the lunisolar perturbation was determined using an analytical model of the orbits of Sun and Moon. The clearest pattern observed in all the simulations was an orbital inclination cycle with a maximum value of 14.5 to 15 degrees and a period of approximately 53 years, which the authors attributed to the precession of the orbital plane discussed above. Kiladze and Sochilina [16] and Kiladze et al. [17] observed the orbital evolution of some 360 uncontrolled GS satellites with the purpose of improving the accuracy of the geopotential model parameters. The results of orbital propagations over intervals of a few years disclosed three types of dynamics, the occurrence of which depended on the initial conditions: (i) a simple libration around the nearest stable position (either about 75º E or about 105º W); (ii) a complex ("long") libration around both stable positions; (iii) a circulation with different drift rates. These behaviors were attributed to the asymmetry of the gravitational field associated to the tesseral and sectorial harmonics of third degree. Changes in the total energy caused by the lunisolar perturbation justified the transitions between regimes of motion. This dynamical behavior was observed in 6 satellites of the set. The orbital analyses carried out by Ariafar and Jehn [18] suggested that retired GS satellites can be placed between 250 and 300 km above the synchronous altitude without them entering a protected 200-km zone over long periods of time (200 years, and probably much longer). Kuznetsov and Kaizer [19] focused on the long-term evolution of geostationary satellites located in the proximity of the separatrices. They identified the regions where the separatrices migrate under the influence of perturbations, and investigated this dynamical behavior making reference to distinct initial inclinations and different surface-to-mass ratios of the spacecraft of interest. More recently, by means of semi-analytical techniques and starting from a wide set of initial conditions (including inclined orbits), Colombo and Gkolias [20] characterized the dynamical structure and long-term stability of the GS orbit region aiming at designing efficient disposal maneuvers.

This work revisits the long-term orbital evolution of uncontrolled GS satellites, improving the accuracy of past studies to gain further understanding of the complex dynamical behavior. High-fidelity numerical propagations include all the relevant orbit perturbations: (i) gravity field synthesis including high-order harmonics for maximum accuracy of the field expansion, (ii) third-body gravitational pull of Sun and Moon, and (iii) effect of solar radiation pressure. The relative positions of Earth, Moon and Sun are determined using ephemerides tables. Orbit propagations use two different sets of equations of motion and numerical integration methods: (i) Gauss planetary equations for modified equinoctial elements with an embedded Runge-Kutta Dormand-Prince 7/8 algorithm; (ii) equations of motion in Cartesian coordinates expressed in an Earth-fixed frame advanced in time with a 13-stage adaptive embedded Runge-Kutta Fehlberg scheme of orders 7 and 8. The results of the orbit propagations with the two methodologies over time intervals of several decades are compared and analyzed. We devote particular attention to the evolution of the satellite's longitude, with the intent of ascertaining if the spacecraft dynamics retains the regular, predictable behavior due to $J_{22}$. Additionally, we propose a geometric approach to describe the precession of the pole of the orbit, and we compare it with the results obtained through numerical propagations.

The reference frames and the physical model adopted in this work are described in Section 2. Section 3 analyzes the precession of the orbital plane, whereas Section 4 describes the two sets of dynamics equations. Section 5 presents the numerical simulations and discusses their results. This is followed (Section 6) by a Fourier analysis of a representative case to illustrate the characteristics and relative importance of the different perturbative effects. Finally, conclusions are drawn in Section 7. A preliminary version of this investigation can be found in [21].

## 2. Physical model

Accurate long-term analysis of geostationary orbits requires accounting for the non-spherical mass distribution of the Earth. A suitable number of spherical harmonics of the geopotential must be included in the calculations in order to obtain realistic results. The third-body gravitational perturbations from Sun and Moon also play an important role. To improve accuracy even more, solar radiation pressure must be modeled as well. Further perturbation sources, like relativistic effects, have not been included because they are several orders of magnitude weaker. This section describes the physical modeling of the relevant forces. As a preliminary step, we define a suitable set of reference frames.

### 2.1 Reference frames

The Earth-Centered quasi-inertial frame (EC) has origin at the Earth's center and is associated with the right-hand sequence of unit vectors $(\hat{i}_1, \hat{i}_2, \hat{i}_3)$, where $\hat{i}_1$ and $\hat{i}_2$ lie in the Earth's mean equatorial plane. In particular, $\hat{i}_1$ is the vernal axis, aligned with the intersection of the mean equatorial and ecliptic planes, while $\hat{i}_3$ points along the Earth rotation axis. This is a satisfactory coordinate system for short-term (and sometimes even medium-term) orbit propagations. However, in the study that follows, long-term orbit propagations (over several decades) are addressed, and the precession of both the Earth equatorial plane and the ecliptic plane must be considered. Due to this secondary motion, the intersection of the two planes is time-varying. A true non-rotating reference frame is obtained by freezing the orientation of the EC axes at a specified time $t_0$ – the initial epoch for orbit propagations in our case. This Earth-Centered Inertial reference system is denoted as $\text{ECI}(t_0)$ henceforth. Its axes $(\hat{c}_1, \hat{c}_2, \hat{c}_3)$ are parallel to the $(\hat{i}_1, \hat{i}_2, \hat{i}_3)$ triad at the reference epoch $t_0$. The transformation from the Geocentric Celestial Reference System (GCRS2000) to the $\text{ECI}(t_0)$ system is governed by the precession relations reported in Appendix. From these expressions, it is possible to derive the rotation matrix $\mathbf{R}_{prec}$ that transforms the $\text{ECI}(t_0)$ frame into the instantaneous mean equator and equinox reference system EC (cf. Appendix),

$$\begin{bmatrix} \hat{i}_1 & \hat{i}_2 & \hat{i}_3 \end{bmatrix}^T = \mathbf{R}_{prec} \begin{bmatrix} \hat{c}_1 & \hat{c}_2 & \hat{c}_3 \end{bmatrix}^T. \tag{1}$$

The motion of the spacecraft is described by its position vector *r* (relative to the Earth's center) and inertial velocity *v*. Two local spacecraft frames are particularly useful for the description of the orbital perturbations:

(a) the local horizontal (LH) frame defined by the triad $(\hat{r}, \hat{E}, \hat{N})$, with $\hat{r}$ parallel to *r*, $\hat{E}$ along the local East direction and $\hat{N}$ aligned with the local North direction;

(b) the local vertical-local horizontal (LVLH) frame defined by the unit vectors $(\hat{r}, \hat{\theta}, \hat{h})$, with $\hat{\theta}$ along the projection of the satellite velocity *v* onto the local horizontal plane and $\hat{h}$ parallel to the specific orbital angular momentum *h*.

The LH-frame is obtained from the EC-frame through a sequence of two elementary rotations: (i) a rotation $\lambda_a$ (absolute longitude) about axis 3 followed by (ii) a negative rotation $\phi$ (latitude) about axis 2,

$$\begin{bmatrix} \hat{r} & \hat{E} & \hat{N} \end{bmatrix}^T = \mathbf{R}_2(-\phi)\mathbf{R}_3(\lambda_a)\begin{bmatrix} \hat{i}_1 & \hat{i}_2 & \hat{i}_3 \end{bmatrix}^T. \tag{2}$$

In Eq. (2), $\mathbf{R}_j(\eta)$ denotes an elementary positive rotation by angle $\eta$ about axis *j*.

The LVLH frame is obtained from the $\text{ECI}(t_0)$ frame through a sequence of three elementary rotations: (i) a rotation $\Omega$ about axis 3, followed by (ii) a rotation $i$ about axis 1, and (iii) a rotation $\theta_t$ about axis 3,

$$\begin{bmatrix} \hat{r} & \hat{\theta} & \hat{h} \end{bmatrix}^T = \mathbf{R}_3(\theta_t)\mathbf{R}_1(i)\mathbf{R}_3(\Omega)\begin{bmatrix} \hat{c}_1 & \hat{c}_2 & \hat{c}_3 \end{bmatrix}^T. \tag{3}$$

In the previous relation, $\Omega$ and $i$ are the right ascension of the ascending node (RAAN) and inclination, respectively. The symbol $\theta_t$ denotes the argument of latitude, defined as $\theta_t := \omega + f$, where $\omega$ and *f* represent the argument of perigee and the true anomaly, respectively.

The coordinates (X, Y, Z) of the satellite's position in the $\text{ECI}(t_0)$ frame are available during propagation (directly in the case of the Cartesian propagator or derived from the osculating orbit elements otherwise). Using Eq. (1), the satellite's position vector

$$\boldsymbol{r} = \begin{bmatrix} X & Y & Z \end{bmatrix} \begin{bmatrix} \hat{c}_1 & \hat{c}_2 & \hat{c}_3 \end{bmatrix}^T \tag{4}$$

can be rewritten as

$$\boldsymbol{r} = \begin{bmatrix} X & Y & Z \end{bmatrix} \mathbf{R}_{prec}^T \begin{bmatrix} \hat{i}_1 & \hat{i}_2 & \hat{i}_3 \end{bmatrix}^T . \tag{5}$$

Since $\boldsymbol{r} = r \begin{bmatrix} \cos\phi\cos\lambda_a & \cos\phi\sin\lambda_a & \sin\phi \end{bmatrix} \begin{bmatrix} \hat{i}_1 & \hat{i}_2 & \hat{i}_3 \end{bmatrix}^T$, Eq. (5) provides $\lambda_a$ and $\phi$.

## 2.2 Geopotential

In the last decades, several accurate gravity field models have been developed. In this investigation, we employ the zero-tide version of Earth Gravitational Model 2008 (EGM2008) [22] which provides the fully-normalized dimensionless Stokes coefficients $C_{nm}$ and $S_{nm}$ and their standard deviations (the magnitude the measurement uncertainty) up to degree ($n$) and order ($m$) 2159. The geopotential $U$ can be expressed as a series of spherical harmonics written in terms of associated Legendre functions of the first kind $P_{nm}$,

$$U = \frac{\mu_E}{r} - \frac{\mu_E}{r}\sum_{n=2}^{\infty}\left(\frac{R_E}{r}\right)^n J_n P_{n0}(\sin\phi) + \frac{\mu_E}{r}\sum_{n=2}^{\infty}\sum_{m=1}^{n}\left(\frac{R_E}{r}\right)^n J_{nm} P_{nm}(\sin\phi)\cos\left[m(\lambda_g - \lambda_{nm})\right], \tag{6}$$

in which $\mu_E$ and $R_E$ are Earth gravitational parameter and equatorial radius, respectively, $\lambda_g$ is the satellite geographical longitude measured from the Greenwich meridian and $J_{nm}$ and $\lambda_{nm}$ relate to $C_{nm}$ and $S_{nm}$ through

$$J_{nm} = \sqrt{C_{nm}^2 + S_{nm}^2}; \quad \lambda_{nm} = \frac{1}{m}\tan^{-1}\left(\frac{S_{nm}}{C_{nm}}\right).$$

If $\theta_G$ denotes Greenwich sidereal time (cf. Appendix), then $\lambda_g = \lambda_a - \theta_G$.

In the LH-frame, the gravitational acceleration is given by

$$\boldsymbol{G} = \nabla U \quad \text{where} \quad \nabla = \hat{r}\frac{\partial}{\partial r} + \frac{\hat{E}}{r\cos\phi}\frac{\partial}{\partial \lambda_g} + \frac{\hat{N}}{r}\frac{\partial}{\partial \phi}. \tag{7}$$

When combined with Eq. (6), this expression yields the three acceleration components $(G_r, G_E, G_N)$. Since the contribution of the main term $\mu_E / r$ is radial, the components of the perturbing acceleration $\boldsymbol{a}^{(H)}$ can be written as $a_r^{(H)} = G_r + \mu/r^2$, $a_E^{(H)} = G_E$, and $a_N^{(H)} = G_N$,

$$\boldsymbol{a}^{(H)} = \begin{bmatrix} a_r^{(H)} & a_E^{(H)} & a_N^{(H)} \end{bmatrix} \begin{bmatrix} \hat{r} & \hat{E} & \hat{N} \end{bmatrix}^T. \tag{8}$$

Equations (1)-(3) allow expressing the components of $\boldsymbol{a}^{(H)}$ in the $\text{ECI}(t_0)$ and LVLH frames.

The previous approach is rather general and yields the instantaneous perturbing acceleration caused by the harmonics of the geopotential. In particular, regarding the $J_2$ term, which expresses the terrestrial polar flattening, simple formulas from analytical averaging are available. The average effect of the $J_2$ perturbation yields no variation of the orbit inclination, semimajor axis, and eccentricity, whereas the orbit plane precesses about the Earth's spin axis with angular velocity [23, 24]

$$\boldsymbol{\omega}_{J_2} = -\frac{3}{2} J_2 \sqrt{\frac{\mu_E}{a^3}} \left(\frac{R_E}{p}\right)^2 (\hat{h} \cdot \hat{i}_3) \hat{i}_3. \tag{9}$$

where $a$ and $p$ are the orbital semimajor axis and semilatus rectum, respectively. Since $\hat{h}$ is parallel to $\hat{i}_3$ if the orbit is equatorial, $J_2$ alone does not alter the motion of a GS satellite (in nominal flight conditions). In the presence of a hypothetical small orbital inclination, the negative sign in Eq. (9) would cause a retrograde precession about $\hat{i}_3$ with a period of 73.5 years. However, the lunisolar gravitational attraction introduces an additional contribution to the precession rate. Section 3 addresses this effect in detail, using a geometric approach.

The perturbation caused by the $J_{22}$ term is particularly relevant due to the synchronism between the spacecraft motion and the Earth rotation. The result is a resonance whose effects can be studied analytically with the Hill-Clohessy-Wiltshire (HCW) equations [24], a first-order approximation of the nonlinear equations of motion of the spacecraft, particularly useful for the purpose of describing the motion of the satellite relative to a reference circular orbit of radius $R_{ref}$,

$$\ddot{x} = 3n^2 x + 2n\dot{y} + a_r^{(J_{22})} \quad \ddot{y} = -2n\dot{x} + a_E^{(J_{22})} \quad \ddot{z} = -n^2 z + a_N^{(J_{22})} \quad \left(n = \sqrt{\mu_E / R_{ref}^3}\right), \tag{10}$$

with $R_{ref} \equiv R_{geo} = 42164$ km, $x = \delta r$, $y = R_{ref}\delta\lambda_a$, and $z = R_{ref}\delta\phi$. The terms $\delta r$, $\delta\lambda_a$, and $\delta\phi$ represent the deviations of $r$, $\lambda_a$ and $\phi$ from their nominal values. If $a_r^{(J_{22})}$, $a_E^{(J_{22})}$, and $a_N^{(J_{22})}$ denote the components of the perturbing acceleration due to $J_{22}$ in the LH-frame, then

$$a_r^{(J_{22})} = -9J_{22}n^2 \frac{R_E^2}{R_{ref}} \cos\left[2\left(\lambda_g - \lambda_{22}\right)\right], \tag{11}$$

$$a_E^{(J_{22})} = -9J_{22}n^2 \frac{R_E^2}{R_{ref}} \sin\left[2\left(\lambda_g - \lambda_{22}\right)\right]. \tag{12}$$

Moreover, $\lambda_g$ may be rewritten as

$$\lambda_g = \lambda_a - \theta_G = \lambda_{a,R} + \delta\lambda_a - \theta_G = \lambda_{g,R} + \frac{y}{R_{ref}}, \tag{13}$$

where $\lambda_{g,R} := \lambda_{a,R} - \theta_G$ is the nominal (and constant) geographical longitude associated with the GS orbit of interest. Under the assumption that $\ddot{x} \simeq 0$ and $\left|3n^2x + 2n\dot{y}\right| \gg \left|a_r^{(J_{22})}\right| \simeq 0$, the first equation in (10) becomes $3n^2x + 2n\dot{y} = 0$, leading to $3n\dot{x} = -2\ddot{y}$. The latter is introduced into the HCW equation for $\ddot{y}$ to yield

$$\ddot{y} = 18J_{22}R_E^2 \frac{n^2}{R_{ref}} \left\{ \sin\left[2\left(\lambda_{g,ref} - \lambda_{22}\right)\right]\cos\left(\frac{2y}{R_{ref}}\right) + \cos\left[2\left(\lambda_{g,ref} - \lambda_{22}\right)\right]\sin\left(\frac{2y}{R_{ref}}\right) \right\}. \tag{14}$$

Figure 1 portrays the behavior of Eq. (14). The radial displacement is exaggerated for the sake of clarity. Oscillatory motion takes place around the two stable equilibrium conditions (denoted with $S_1$ and $S_2$), with amplitude and period completely determined by the initial conditions. Two unstable equilibrium points exist at geographic longitudes -14.9° ($I_1$) and 165.1° ($I_2$). When further perturbations are considered, the long-term orbital evolution becomes complex, especially if the initial position of the satellite is close to one of the two unstable equilibrium points. Moreover, the two assumptions $\ddot{x} \simeq 0$ and $a_r^{(J_{22})} \simeq 0$ can be checked a posteriori, by comparing the analytical results with the numerical propagations. These issues are investigated in the sections that follow.

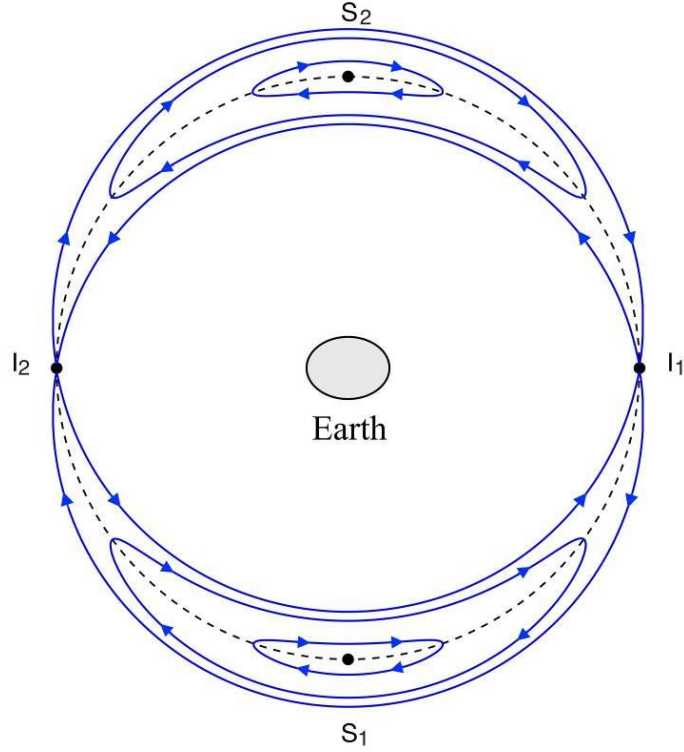

Figure 1: Longitudinal motion due to $J_{22}$. Dashed black line: nominal GS orbit,

$S_1$ and $S_2$: stable points, $I_1$ and $I_2$: unstable points.

## 2.3 Third-body perturbation

The gravitational pull of Moon and Sun has significant effects on GS orbits. This third-body perturbation is expressed as the differential action exerted on the satellite and on Earth. The perturbing acceleration due to a third body is

$$\boldsymbol{a}^{(M/S)} = -\frac{\mu_3}{s_3^3 (1+q_3)^{3/2}} \left[ \boldsymbol{r} + s_3 q_3 \frac{3 + 3q_3 + q_3^2}{1 + (1+q_3)^{3/2}} \right] \quad \text{where} \quad q_3 := \frac{r^2 - 2\boldsymbol{r} \cdot \boldsymbol{s}_3}{s_3^2}. \tag{15}$$

Superscript (M/S) refers to either the Moon (M) or the Sun (S), whereas the symbol $\mu_3$ denotes the gravitational parameter of the third body, $\boldsymbol{s}_3$ is its position vector relative to the Earth, and $s_3 = |\boldsymbol{s}_3|$. The previous expression makes use of the Battin-Giorgi [25, 26] approach to Encke's method for special perturbations. The position vector $\boldsymbol{s}_3$ is expressed in the $\text{ECI}(t_0)$ frame using planetary ephemerides (cf. Appendix). Then, the components of $\boldsymbol{s}_3$ in the LVLH frame can be obtained by means of Eq. (3). The third-body perturbing acceleration $\boldsymbol{a}^{(3B)}$ is the sum of the contributions from Moon and Sun, i.e. $\boldsymbol{a}^{(3B)} = \boldsymbol{a}^{(M)} + \boldsymbol{a}^{(S)}$.

Under the assumptions that the spacecraft moves in a circular orbit of radius $R_{ref}\ (=R_{geo})$ and that the third body describes a circular orbit of radius $r_{12}$ relative to the Earth, the double averaging of Gauss equations for perturbed orbits yields the precession rate of the orbital plane [17]

$$\boldsymbol{\omega}_{3B}^{(M/S)} = -\frac{3}{4}\frac{\mu_3 R_{ref}^2}{h r_{12}^3}\left(\hat{h}\cdot\hat{h}_{3B}\right)\hat{h}_{3B}, \qquad (16)$$

in which $\hat{h}_{3B}$ is the unit vector normal to the plane of motion of the third body relative to the Earth and $h = |\boldsymbol{h}|$.

## 2.4 Solar radiation pressure

In this work, the perturbing acceleration due to the solar radiation pressure is described using the so-called cannonball model [23],

$$\boldsymbol{a}^{(SR)} = -\upsilon P_{SR}\frac{c_R S_R}{\tilde{m}}\hat{r}_S, \qquad (17)$$

where $P_{SR}\ (=4.557\cdot 10^{-6}\ \text{Pa})$ is the solar radiation pressure at 1 astronomical unit from the Sun, $c_R$ is the radiative coefficient of the satellite's surface, $\tilde{m}$ and $S_R$ are the satellite's mass and cross section, respectively, $\boldsymbol{r}_S$ is the position of the Sun relative to the Earth, $\hat{r}_S = \boldsymbol{r}_S/|\boldsymbol{r}_S|$, and $\upsilon$ is a shadow function. In the $\text{ECI}(t_0)$ frame, $\boldsymbol{r}_S$ is obtained using planetary ephemerides (cf. Appendix). Then, $\boldsymbol{r}_S$ can be projected onto the LVLH-frame using Eq. (3) to yield the three components of the perturbing acceleration $\left(a_r^{(SR)}, a_\theta^{(SR)}, a_h^{(SR)}\right)$.

The shadow function $\upsilon$ equals either 0 (when the space vehicle is eclipsed) or 1 (when it is illuminated). Earth's shadow is assumed cylindrical. Thus, letting $\vartheta_1 := \arccos(R_E/r)$ and $\varphi := \arccos(\hat{r}\bullet\hat{r}_S)$, the space vehicle is eclipsed if [23]

$$\varphi > \vartheta_1 + \frac{\pi}{2}. \qquad (18)$$

When inequality (18) holds, $\upsilon = 0$, otherwise $\upsilon = 1$.

## 3. Geometric analysis of the orbit precession

This section presents an approximate geometric analysis of the precessional motion of the orbital plane resulting from the combined action of $J_2$ and the third-body perturbation of Moon and Sun. This analysis is based on Eqs. (9) and (16) (cf. also Ref. 17, 23, and 24). The overall angular rate of precession is $\boldsymbol{\omega}_{pr} = \boldsymbol{\omega}_{J2} + \boldsymbol{\omega}_{3B}^{(M)} + \boldsymbol{\omega}_{3B}^{(S)}$. Two simplifying assumptions are introduced: (i) $\hat{i}_3 \simeq \hat{c}_3$ – the precession of the ecliptic and equatorial planes is neglected – and (ii) the angular orbital momentum of the Moon is aligned with the ecliptic pole. These assumptions imply that $(-\boldsymbol{\omega}_{pr})$ is displaced by an angle $\delta = 7.1°$ from $\hat{c}_3$, and the angular momentum describes (in clockwise sense) the cone portrayed in Fig. 2(a). Therefore, using Eqs. (9) and (16), the (constant) precession rate $\omega_{pr}$ can be evaluated, and the precession period $T_{pr}$ equals 52 years $\left(T_{pr} = 2\pi/\omega_{pr}; \; \omega_{pr} = |\boldsymbol{\omega}_{pr}|\right)$. The right-handed sequence $(\hat{c}_1, \hat{s}, \hat{p})$ of Fig. 2(b) is obtained from $(\hat{c}_1, \hat{c}_2, \hat{c}_3)$ through a rotation by angle $\delta$ about axis 1,

$$\begin{bmatrix} \hat{c}_1 & \hat{s} & \hat{p} \end{bmatrix}^T = \mathbf{R}_1(\delta) \begin{bmatrix} \hat{c}_1 & \hat{c}_2 & \hat{c}_3 \end{bmatrix}^T. \tag{19}$$

In Fig. 2(b), angles $\delta$ and $\chi$ identify the instantaneous angular momentum $\boldsymbol{h}$ in the $(\hat{c}_1, \hat{s}, \hat{p})$ frame. Then, $\boldsymbol{h}$ can be projected in the $(\hat{c}_1, \hat{c}_2, \hat{c}_3)$ frame using Eq. (19),

$$\boldsymbol{h} = h[\sin\delta\cos\chi \;\; \sin\delta\cos\chi \;\; \cos\delta] \begin{bmatrix} \hat{c}_1 & \hat{s} & \hat{p} \end{bmatrix}^T = h[\sin\delta\cos\chi \;\; \sin\delta\cos\chi \;\; \cos\delta]\mathbf{R}_1(\delta)\begin{bmatrix} \hat{c}_1 & \hat{c}_2 & \hat{c}_3 \end{bmatrix}^T. \tag{20}$$

The components of $\boldsymbol{h}$ in the $(\hat{c}_1, \hat{c}_2, \hat{c}_3)$ frame can be written also in terms of $\Omega$ and $i$,

$$\boldsymbol{h} = h[\sin i \sin\Omega \;\; -\sin i \cos\Omega \;\; \cos i]\begin{bmatrix} \hat{c}_1 & \hat{c}_2 & \hat{c}_3 \end{bmatrix}^T \tag{21}$$

Comparison of Eqs. (20) and (21) yields the following relations:

$$i = \arccos\left(\sin^2\delta\cos\chi + \cos^2\delta\right), \tag{22}$$

$$\sin\Omega = \frac{\sin\delta\sin\chi}{\sin i} \quad \text{and} \quad \cos\Omega = \frac{\sin\delta\cos\delta(1-\cos\chi)}{\sin i}. \tag{23}$$

Because $0 \le \delta \le \pi/2$, Eq. (22) gives

$$\sin i = \sqrt{1 - \left(s_\delta^2 c_\chi + c_\delta^2\right)^2} = s_\delta\sqrt{1-c_\chi}\sqrt{1+s_\delta^2 c_\chi + c_\delta^2} = s_\delta\sqrt{2}\left|\sin\frac{\chi}{2}\right|\sqrt{1+s_\delta^2 c_\chi + c_\delta^2}, \tag{24}$$

where $s_\eta := \sin\eta$ and $c_\eta := \cos\eta$ ($\eta$ denotes a generic angle). Substitution of Eq. (24) into Eq. (23) leads to

$$\sin\Omega = \sqrt{2}\,\frac{\cos\frac{\chi}{2}\sin\frac{\chi}{2}}{\left|\sin\frac{\chi}{2}\right|\sqrt{1+s_\delta^2 c_\chi + c_\delta^2}} \quad \text{and} \quad \cos\Omega = \sqrt{2}\,\frac{c_\delta\left|\sin\frac{\chi}{2}\right|}{\sqrt{1+s_\delta^2 c_\chi + c_\delta^2}} \quad \Rightarrow \quad \tan\Omega = \frac{c_\delta}{\tan\frac{\chi}{2}}. \tag{25}$$

The previous relations prove that $\Omega$ varies with the same period as $\chi$ and

$$\Omega \to \frac{\pi}{2} \text{ as } \chi \to 0^+ \quad \text{and} \quad \Omega \to -\frac{\pi}{2} \text{ as } \chi \to 2\pi^- \tag{26}$$

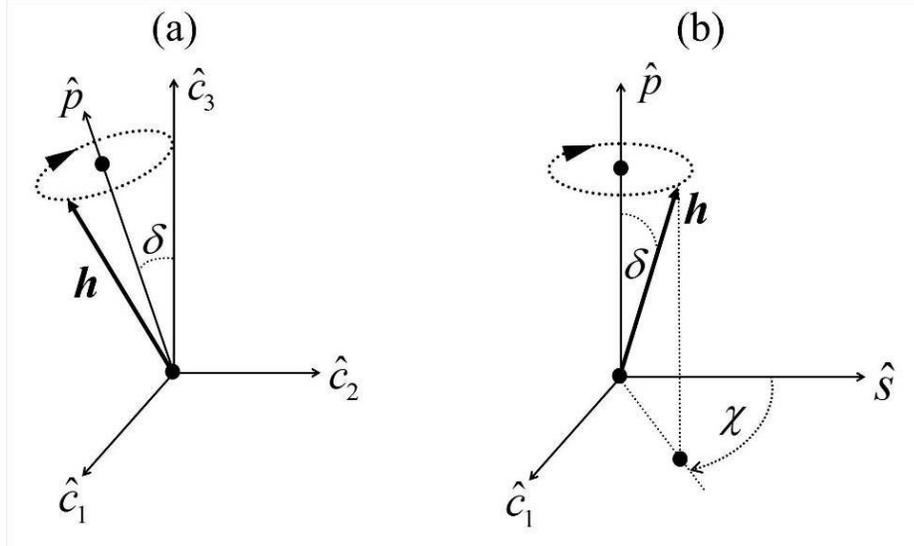

Figure 2: Geometric illustration of the precessional motion.

Moreover, since $\delta \approx 7.1$ deg, $\tan\Omega\tan(\chi/2)\approx 1$ and

$$\tan\frac{\chi}{2}\tan\Omega = 1 - \frac{\tan\frac{\chi}{2}+\tan\Omega}{\tan\left(\Omega+\frac{\chi}{2}\right)} \approx 1 \quad \Rightarrow \quad \frac{\tan\frac{\chi}{2}+\tan\Omega}{\tan\left(\Omega+\frac{\chi}{2}\right)} \approx 0. \tag{27}$$

The last relation yields two possible (approximate) solutions: (i) $\tan(\chi/2)+\tan\Omega \approx 0$ and (ii) $\Omega+\chi/2 \approx \pi/2$. However, option (i) is not compatible with $\tan\Omega\tan(\chi/2)\approx 1$, therefore

$$\Omega \approx \frac{\pi}{2}-\frac{\chi}{2}, \quad \text{with } 0 \le \chi \le 2\pi. \tag{28}$$

The previous relation refers to a specific interval for $\chi$. However, $\Omega$ is periodic, thus the previous expression of $\Omega$ as a function of $\chi$ can be easily extended to subsequent precession periods (for $\chi > 2\pi$). Finally, due to the small value of $\delta$, Eq. (24) yields

$$i \approx 2\delta \left|\sin\frac{\chi}{2}\right|. \tag{29}$$

Equations (28)-(29) represent two simple approximate relations. They will be compared to the results of accurate numerical simulations in Section 5.1.

## 4. Orbit Dynamics

The satellite orbit evolution can be described in terms of either spherical coordinates, Cartesian coordinates, or osculating orbit elements, i.e. semimajor axis $a$, eccentricity $e$, inclination $i$, RAAN $\Omega$, argument of periapse $\omega$, and true anomaly $f$ [24]. This section presents two distinct equation sets tailored to describing the long-term orbit evolution of decommissioned geostationary satellites, in the presence of the perturbations treated in Section 2.

### 4.1 Modified equinoctial elements

Gauss planetary equations become singular for circular and equatorial orbits, and also when an elliptical orbit transitions to a hyperbola. The modified equinoctial orbital elements [27]

$$p = a(1-e^2) \quad l = e\cos(\Omega+\omega) \quad m = e\sin(\Omega+\omega) \quad n = \tan\frac{i}{2}\cos\Omega \quad s = \tan\frac{i}{2}\sin\Omega \quad q = \Omega+\omega+f \tag{30}$$

eliminate the singularities except for the case of equatorial retrograde orbits ($i = 180°$). The equations that govern the variation of the equinoctial elements are [25, 28]

$$\dot{p} = \frac{2}{\vartheta}\sqrt{\frac{p^3}{\mu_E}}a_\theta, \tag{31}$$

$$\dot{l} = \sqrt{\frac{p}{\mu_E}}\left[a_r\sin q + a_\theta\frac{(\vartheta+1)\cos q + l}{\vartheta} - a_h\frac{n\sin q - s\cos q}{\vartheta}m\right], \tag{32}$$

$$\dot{m} = \sqrt{\frac{p}{\mu_E}}\left[-a_r\cos q + a_\theta\frac{(\vartheta+1)\sin q + m}{\vartheta} + a_h\frac{n\sin q - s\cos q}{\vartheta}l\right], \tag{33}$$

$$\dot{n} = a_h \sqrt{\frac{p}{\mu_E}} \frac{1+n^2+s^2}{2\vartheta} \cos q, \tag{34}$$

$$\dot{s} = a_h \sqrt{\frac{p}{\mu_E}} \frac{1+n^2+s^2}{2\vartheta} \sin q, \tag{35}$$

$$\dot{q} = \sqrt{\frac{\mu_E}{p^3}} \vartheta^2 + a_h \sqrt{\frac{p}{\mu_E}} \frac{n \sin q - s \cos q}{\vartheta}, \tag{36}$$

in which $\vartheta = 1 + l\cos q + m\sin q$ and $r = p/\vartheta$. The terms $\{a_r, a_\theta, a_h\}$ are the components of the perturbing acceleration $a$ in the LVLH-frame, i.e., $a = a^{(H)} + a^{(3B)} + a^{(SR)}$.

## 4.2 Cartesian coordinates

The Cartesian propagator solves the dynamics of the satellite in a TRS (Terrestrial Reference System) attached to Earth (i.e., the planet remains stationary in the reference frame) and aligned with the right-hand sequence of unit vectors $(\hat{i}, \hat{j}, \hat{k})$. TRS is obtained rotating the EC frame by the angle $\theta_G$ about its third axis. This is convenient for computing Earth's gravitational acceleration and keeps the satellite velocity small at all times, making the numerical integration more robust.

Given that TRS rotates with respect to the fixed stars, inertial forces must be included in the computation, therefore $a^{TRS} = G + a^{(3B)} + a^{(SR)} + a^{(I)}$. The inertial acceleration is

$$a^{(I)} = -\Omega^{TRS} \times (\Omega^{TRS} \times r) - 2\Omega^{TRS} \times \dot{r}, \tag{37}$$

where $\Omega^{TRS}$ denotes the angular velocity of TRS respect to ECI. This angular velocity can be written as $\Omega^{TRS} = \Omega^{sidereal} + \Omega^{EC}$. The first term of the sum is Earth's sidereal rate – the angular velocity of TRS respect to EC – and the second is the angular velocity of precession of the EC frame. In practice, $\Omega^{EC}$ is so small that it can be neglected in the computation of inertial forces. The ODE governing the motion of the satellite is

$$\frac{d}{dt}\begin{bmatrix} r \\ \dot{r} \end{bmatrix} = \begin{bmatrix} \dot{r} \\ a^{TRS} \end{bmatrix}, \tag{38}$$

where the position and velocity vectors are expressed in TRS. The Cartesian propagator includes two different schemes for computing the gravitational acceleration. The first one uses the traditional formulation in spherical coordinates

$$U(r, \lambda_g, \phi) = \frac{\mu_E}{r}\left(1 + \sum_{n=2}^{N}\left(\frac{R_E}{r}\right)^n \Omega_n\right), \tag{39}$$

where the partial sums of degree $n$

$$\Omega_n = \sum_{m=0}^{n} P_{nm}(\sin\phi)\left(C_{nm}\cos m\lambda_g + S_{nm}\sin m\lambda_g\right) \tag{40}$$

are computed with the forward row recursion algorithm of [29]. It is computationally very efficient and suitable for ultra-high expansion degrees ($N>2000$) in geodesy applications. It is not directly applicable to polar orbits, however, due to the singularity associated with the spherical coordinate system. For arbitrary orbits, the propagator includes a harmonic synthesis module operating directly in Cartesian coordinates. The algorithm [30] replaces the associated Legendre functions with Helmholtz polynomials $H_{nm}$ [31]. The geopotential is rewritten in terms of the direction cosines of the position vector

$$s = \frac{X}{r}\ ;\ t = \frac{Y}{r}\ ;\ u = \frac{Z}{r}, \tag{41}$$

yielding a series of the form

$$U(s,t,u,r) = \sum_{n=0}^{N}\rho_n \sum_{m=0}^{n} D_{nm}(s,t) H_n^m(u). \tag{42}$$

The term $\rho_n$ in Eq. (42) is named the parallactic factor

$$\rho_n = \frac{\mu_E}{r}\left(\frac{R_E}{r}\right)^n, \tag{43}$$

while $D_{nm}$ is the mass coefficient of degree $n$ and order $m$

$$D_{nm}(s,t) = C_{nm}\operatorname{Re}\left((s+it)^m\right) - S_{nm}\operatorname{Im}\left((s+it)^m\right). \tag{44}$$

Details on the algorithm as well as efficient recursion schemes for computing the sums can be found in [32]. Both harmonic synthesis codes (i.e., in spherical/Cartesian coordinates) agree to within one part in $10^{14}$ when operating in double precision − 64-bit floating point representation − arithmetic. Their accuracy has also been verified through comparisons with other popular schemes [32].

## 5. Long-term orbit propagations

In this section, the long-term orbit evolution of decommissioned GS satellites is investigated using the two sets of equations of motion introduced in Section 4. In both cases, the perturbations included in the physical model are the terrestrial gravitational harmonics up to degree and order 8, the third-body pull of Moon and Sun and the solar radiation pressure.

The 8$^{th}$-degree expansion was selected because it delivers the best computational performance while retaining the full accuracy of the EGM2008 model. Using the central limit theorem and the standard deviation of the Stokes coefficients, the standard deviation of the gravitational acceleration at any point in space can be determined [33]. Sampling the standard deviation over a collection of points of a sphere, the typical uncertainty of the model at a given height can be estimated. Then, the typical value of the truncation error can be evaluated at the same sampling points. Equating the uncertainty of the geopotential model to the truncation error, the lowest expansion degree that delivers the maximum accuracy is determined. For a GS orbit this degree is 8, corresponding to an acceleration uncertainty of $3 \cdot 10^{-13}$ m/s$^2$. Adding spherical harmonics beyond this degree would not improve the precision, because the intrinsic uncertainty of the EGM2008 model would dominate over the truncation error of the series.

For the computation of the perturbation caused by the solar radiation, a mass $\tilde{m}=3000$ kg, a cross section $S_R=10$ m$^2$ and a radiative coefficient $c_R=2$ (perfect reflection) have been assumed.

### 5.1 Numerical integration methods and settings

Orbit propagations have been carried out using the following two approaches:

(A) Integration of the Gauss equations (31)-(36), in conjunction with the Matlab routine *ode87* [34], which uses an explicit Runge-Kutta method based on 8-7$^{th}$-order Dorman and Prince formulas [35,36]. This numerical integration algorithm requires 13 function evaluations per integration step. Relative accuracies of orders $10^{-14}$ and $10^{-15}$ were used. Moreover, canonical units were employed for the variables of interest, i.e. the distance unit (DU), set to the value of the Earth radius, and the time unit (TU), such that $\mu_E = 1$ DU$^3$/TU$^2$.

(B) Integration of the Cartesian equations (38). The propagator implements a variety of explicit Runge-Kutta integration schemes. Upon testing methods with orders from 3 to 9, the best computational performance was achieved with a 13-stage adaptive embedded scheme of orders 7 and 8 due to Fehlberg [37]. Linear extrapolation was used, retaining the 8th order solution. For each case (combination of initial longitude and date) the propagation has been repeated with relative integrator tolerances $10^{-11}$, $10^{-12}$, $10^{-13}$, $10^{-14}$ and $10^{-15}$ to investigate the convergence of the solution and assess the reliability of the results.

The adoption of two completely different propagation codes gives additional confidence in the results. For each tested case, the orbit propagation is considered reliable if the results from the two methods agree over sufficiently long time intervals.

Two initial epochs ($t_0$) have been used: January 1st 2020 at 0:00 UTC (JDN 2458849.5) and June 1st 2030 at 0:00 UTC (JDN 2462653.5) to incorporate the effect of changes in geometry of the Earth-Sun-Moon system. The initial radius and latitude of the satellite are set to $r = R_{GEO}$, $\phi = 0°$. 64 equally-spaced values of the initial geographical longitude $\lambda_g$ were tested for each initial date. The values of $\lambda_g$ lie inside the intervals [150°,180°] and [-30°,0°], which are centered on the unstable equilibrium points $I_2$ and $I_1$. The proximity of the initial position to the unstable equilibrium points increases the likelihood of complex longitudinal behavior. All combinations of initial longitude and epoch have been propagated for at least 150 years. A subset of scenarios has been computed for 1000 years to gain further insight into the long-term behavior.

## 5.2 Precession motion

As shown in Section 3, the out-of-plane satellite dynamics is dominated by the third-body perturbation combined with the effect of the $J_2$ term. Figures 3 and 4 show the long-term variations of inclination and RAAN. The two orbit elements exhibit the same repeating patterns regardless of the initial longitude. The analytical approximate model and the related relations yield predictions in excellent agreement with the numerical propagations over several decades.

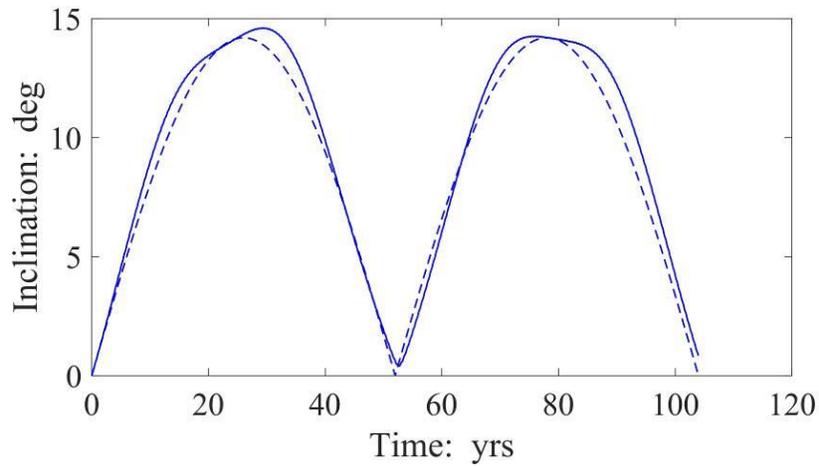

Figure 3: Time history of the inclination. Initial epoch: January 1st 2020, initial geographical longitude: 156º. Dashed line: analytical approximation; solid line: numerical propagation.

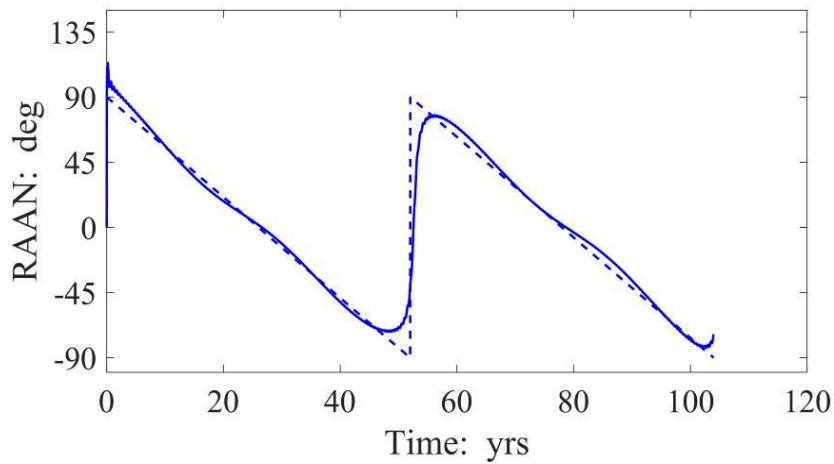

Figure 4: Time history of the RAAN. Initial epoch: January 1st 2020, initial geographical longitude: 156º. Dashed line: analytical approximation, solid line: numerical propagation.

### 5.3 Longitudinal dynamics

Unlike RAAN and inclination, the longitudinal evolution can be very complex. To better understand the underlying mechanisms, it is easier to start considering the effect of Earth's gravity alone. Figures 5 depicts the variation of the gravitational potential along the geostationary ring. The potential has been calculated with expansion degrees 2, 3 and 8 to illustrate the effect of the different harmonics. If only 2nd degree harmonics are included, the potential is symmetric respect to $I_1$. In this case, the dynamics agree very well with the simple $J_{22}$ theory of Section 2.2. A satellite initially at rest in the TRS frame will drift toward the closest stable equilibrium point and orbit around it (motion types $T1$ and $T2$ in the figure). Note that, counterintuitively at first glance, the longitudinal component of the gravitational force is repulsive at the stable equilibrium points, while it is attractive at the unstable ones. This behavior is caused by the rotating character of the TRS frame. If a satellite drifts slightly forward – i.e., eastward – of a stable equilibrium point, the gravitational acceleration will push it further East. This increases the energy of the satellite, raising it to a higher orbit. The orbital period then increases, allowing Earth's rotation to catch up with the satellite, which brings it back towards the equilibrium point. Conversely, if the spacecraft drifts westward of the stable equilibrium point, gravity will push it backward lowering the orbit. This will reduce the orbital period allowing the satellite to overtake the Earth's rotation. In general, the satellite will drift eastward whenever the semimajor axis falls below the geostationary radius, and westward when it is larger.

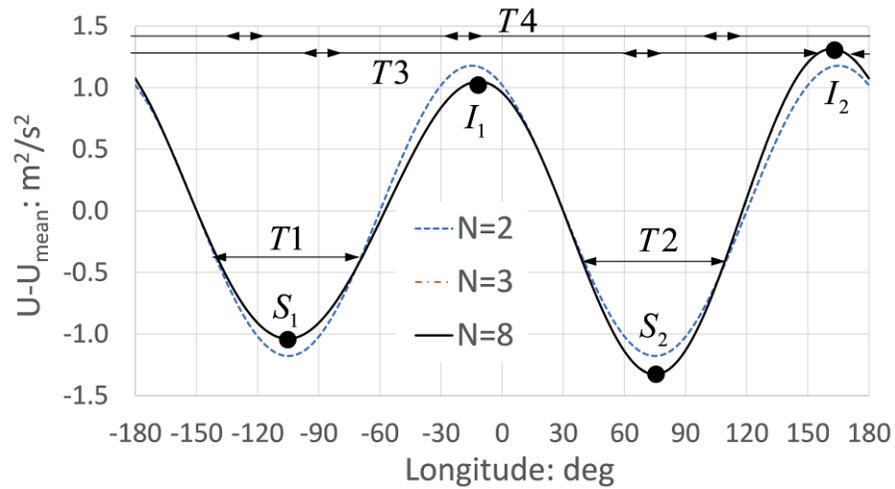

Figure 5: Longitudinal variation of the potential for field expansion degrees 2, 3 and 8. Unstable ($I_1$, $I_2$) and stable ($S_1$, $S_2$) equilibrium points. Typical satellite drift patterns ($T1$-$T4$).

Once 3$^{rd}$ degree harmonics are included, the symmetry of the geopotential is broken. Both the position of the equilibrium points and the magnitude of the extrema change. Figures 5 shows that the curves for N=3 and 8 are almost coincident. Adding harmonics above third degree has an extremely weak effect on the potential. The positions of the unstable equilibrium points – when high-degree harmonics are considered – are

$$\lambda_g(I_1) = -11.5° \quad \text{and} \quad \lambda_g(I_2) = 161.9°. \tag{45}$$

As the heights of the two peaks are different, a new type of motion becomes possible. A satellite that starts close to $I_2$ has enough energy to travel across $I_1$. This results in a wide oscillation spanning most longitudes (T3) except for a small exclusion zone centered on the second unstable point.

Adding the lunisolar perturbation introduces a subtle modulation of the geopotential curve [16] with a period close to 52 years, as outlined in section 3. In the numerical propagations this variation is not truly periodic, because the third-body positions are derived from ephemerides instead of simple analytical models. Both the position and the height of the peaks are affected. When the alignment of the Moon and Sun is favorable, the change in height of the extrema allows passages across an unstable equilibrium point that would not be possible otherwise. For example, a satellite that starts with motion type *T*1 might transition to mode *T*3 for a limited time and then back to *T*1 or *T*2. A transition between *T*3 and *T*4 − continuous circulation across all longitudes − is also possible. In *T*4 mode the longitude evolution is monotonic − i.e., drift with constant East/West direction − until a transition to *T*3 mode happens. The energy difference between the two unstable points is so large that transitions between states *T*1/*T*2 and *T*4 were not encountered, even for propagations spanning one millennium.

The effect of the solar radiation pressure is minor compared to the lunisolar tidal forces, so it does not introduce any new behavior. It can still affect the timing of transitions, which are extremely sensitive to small perturbations (more details on this are provided in the next paragraph).

The transitions, when they occur, do not follow any simple pattern, so the only way of predicting them is to use numerical propagation. We shall distinguish two overall behaviors: regular drift (R) with no transitions − i.e, pure *T*1, *T*2 or *T*3 motion − and complex aperiodic motion (C) involving transitions between different drift patterns.

R-behavior can be predicted accurately over very long time intervals (more than 1000 years). Irregular (C) motion, on the other hand, can only be propagated reliably over limited periods. For motion (C), there are always dates into the future when the satellite comes to a near-stop at an unstable point where all the forces nearly cancel out. Let us call this an "instantaneous unstable equilibrium point" (IUEP) in the sense that it changes continuously, because it depends on the positions of Sun and Moon. It is worth remarking that it is not a true equilibrium point, but a combination of time and position. For reference, the list that follows gives the position of the IUEPs at the two initial epochs.

$$\text{January 1}^{\text{st}} \text{ 2020 at 0:00 UTC}: \quad \lambda_g\left(\text{IUEP}_1\right) = -13.5° \quad \text{and} \quad \lambda_g\left(\text{IUEP}_2\right) = 160.2°$$
$$\text{June 1}^{\text{st}} \text{ 2030 at 0:00 UTC}: \quad \lambda_g\left(\text{IUEP}_1\right) = -16.0° \quad \text{and} \quad \lambda_g\left(\text{IUEP}_2\right) = 157.8°$$
(46)

Whether or not the satellite moves across the IUEP − i.e., the occurrence of a transition − is extremely sensitive to spurious perturbations such as numerical errors. This makes reliable predictions of the occurrence of a transition at the IUEP virtually impossible, because discretization and rounding errors cannot be completely eliminated. For well-behaved cases, when the integrator tolerance is decreased, the results agree over increasingly long periods of time. Robust predictions for more than 150 years are possible in most cases, with the majority remaining consistent for over 3 centuries. There is a small subset of cases where the critical condition − satellite at rest at an IUEP − occurs early during the simulation. For these "pathological" scenarios, reducing the tolerance does not improve the predictions. The smaller time steps increase the accumulation of rounding errors and these, due to the extreme sensitivity of the solution, contaminate the results. Given that the rounding errors depend on the exact sequence of operations each algorithm uses, the results from the two codes (Equinoctial elements/Cartesian coordinates) disagree after the critical point. Using quad-precision arithmetic − 128-bit floating point representation − can delay, but not avoid, the onset of unpredictability. The unpredictability refers only to whether or not the satellite moves across the IUEP. However, the general pattern of motion does not change.

List (47) details the three worst-behaved cases in our set, including the duration of the agreement between the two propagators ($\Delta t$). For the 61 remaining scenarios the results coincide over 150 years or more.

(1) $\Delta t = 88$ years: initial epoch January 1$^{\text{st}}$ 2020 0:00 UTC, $\lambda_g(t_0) = 156°$
(2) $\Delta t = 58$ years: initial epoch January 1$^{\text{st}}$ 2020 0:00 UTC, $\lambda_g(t_0) = -10°$  (47)
(3) $\Delta t = 57$ years: initial epoch June 1$^{\text{st}}$ 2030 0:00 UTC, $\lambda_g(t_0) = 158°$

Tables 1 and 2 summarize the major findings related to the longitudinal motion, including the class of behavior (R or C), the types of motion (*T*1/2/3/4) and the longitude range covered ($\lambda_g^{(min)}$ to $\lambda_g^{(max)}$) if the satellite does not complete a full revolution.

An example of R-behavior (orbit type *T*1) is shown in Figs. 6-7. This evolution is predicted with reasonable accuracy using the analytical model Eq. (14). Figs. 8-9 portray another regular evolution, this time around the second stable equilibrium point (type *T*2). The figures illustrate clearly the aforementioned correlation between semimajor axis and direction of drift.

A complex (C) scenario with irregular transitions between the *T*3 and *T*4 states is charted in Figs. 10 and 11. An interesting variation is depicted in Figs. 12-13. This seems, at first glance, a regular *T*4 behavior. Under appropriate initial conditions, the satellite can be trapped in a *T*4 state for very long periods, but not permanently. This occurs when the starting point is close to the second IUEP at an epoch when the potential is near its historic maximum (i.e., the highest value over the 52-year cycle). This means the spacecraft starts moving with a large energy. By the next time it reaches $I_2$, the height of the geopotential peak will be lower due to the change in position of Sun and Moon. Therefore, the energy of the satellite allows it to move across the unstable point. Given that the combination of position of the satellite and third bodies that leads to the high energy state happens very infrequently, the transition between *T*4 and *T*3 becomes exceedingly rare. The motion might seem pure mode *T*4 if the duration of the propagation is insufficient. This is the case for Figs. 12-13, where a transition to mode *T*3 for just one cycle occurs at 155 years; the satellite then reverts to mode *T*4. In more extreme cases, the system can stay locked in mode *T*4 for several centuries. For example, the scenario starting in 2030 with $\lambda_g = 160°$ remains in mode *T*4 for the first 580 years, transitions to *T*3 for a single cycle and immediately goes back to *T*4. These fringe cases, although (C) *T*3-*T*4 in strict terms, could in practice be considered regular *T*4 motion with very good approximation. They have been labelled R* 4* in table 2 to highlight this fact. The scenarios with initial date 2030 and $\lambda_g \in \left[160°, 164°\right]$ belong to this category.

Finally, Figs. 14-17 illustrate the first of the ill-behaved scenarios from list (47). Figs. 14 and 15 show results from the equinoctial elements propagator while data in Figs. 16 and 17 come from the Cartesian integrator. While the curves overlap just for the first 88 years, the discrepancy is due only to different predictions for the time of a transition between regimes *T*4 and *T*3. The overall motion patterns predicted by the two methods remain qualitatively the same. Therefore, the differences are not relevant for the characterization of the satellite evolution.

In all 64 scenarios tested, the out-of-plane motion − the evolution of inclination and RAAN − is very similar to that shown in Figs. 3 and 4. Hence, it has not been reported to reduce clutter. It is worth remarking that the existence of transitions between different behaviors was described by Kuznetsov and Kaiser [19] in terms of migration of the separatrices. The present research identifies and classifies these transitions through orbit propagations over several decades (or even centuries), and clarifies their relation with the time-varying effect of perturbations (mainly, higher-order harmonics of the geopotential and the lunisolar gravitational attraction).

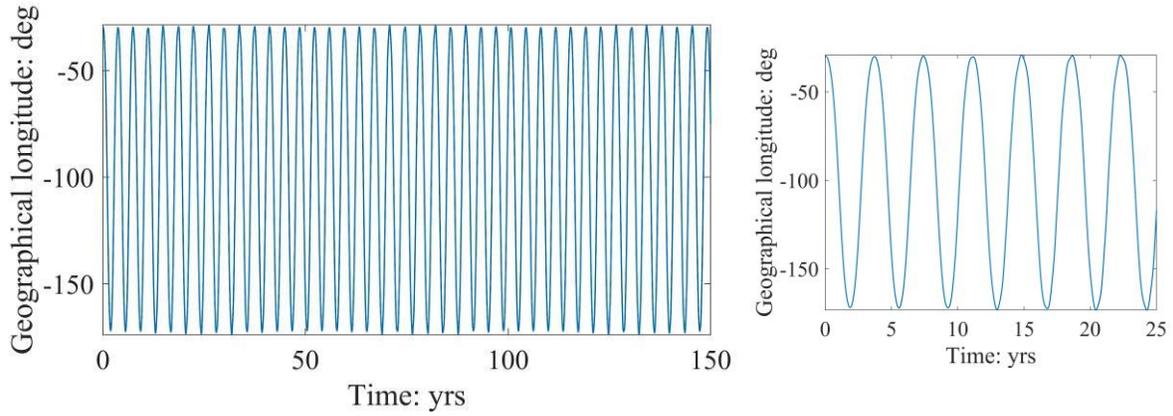

Figure 6: Time history of the geographical longitude. Detail of the first 25 years on the right.

Initial epoch: January 1$^{st}$ 2020, initial longitude: -30º.

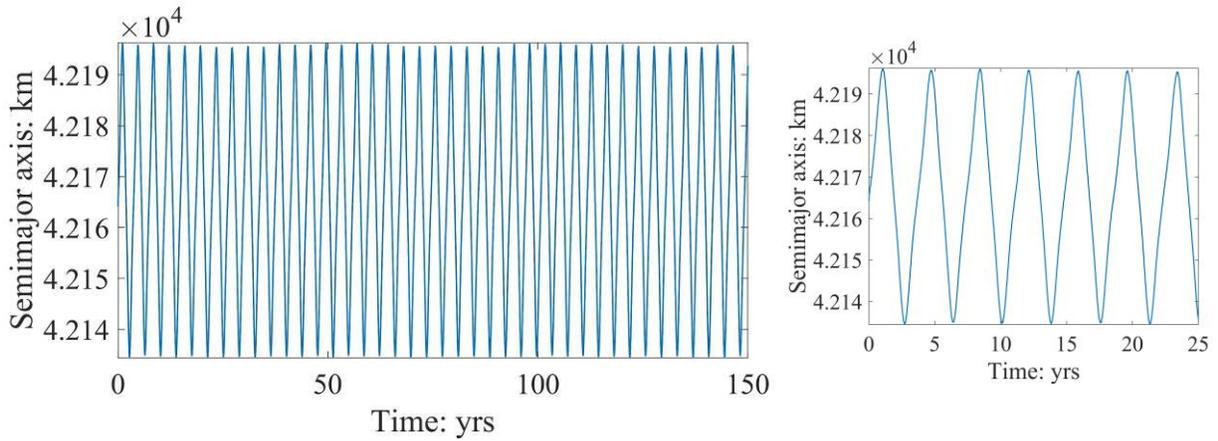

Figure 7: Time history of the semimajor axis. Detail of the first 25 years on the right.

Initial epoch: January 1$^{st}$ 2020, initial longitude: -30º.

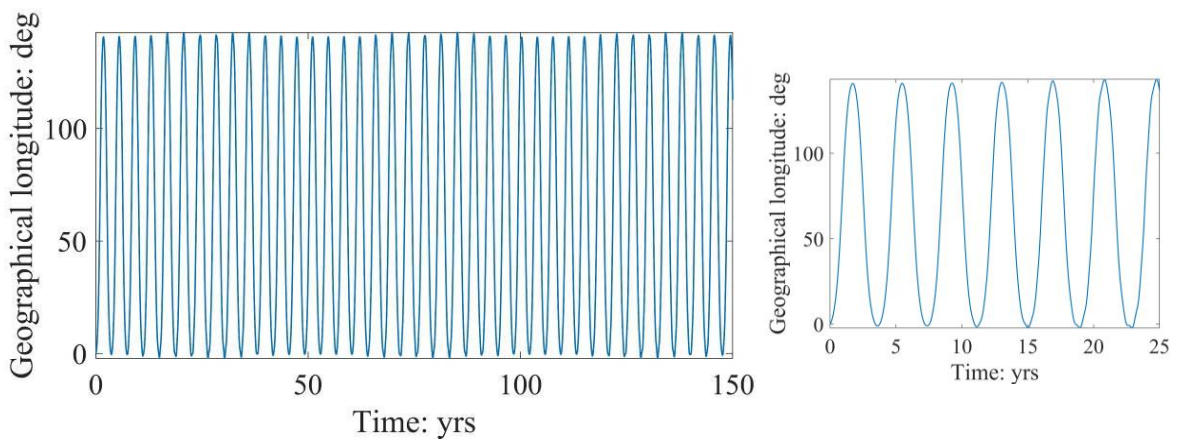

Figure 8: Time history of the geographical longitude. Detail of the first 25 years on the right.

Initial epoch: June 1$^{st}$ 2030, initial longitude: 0º.

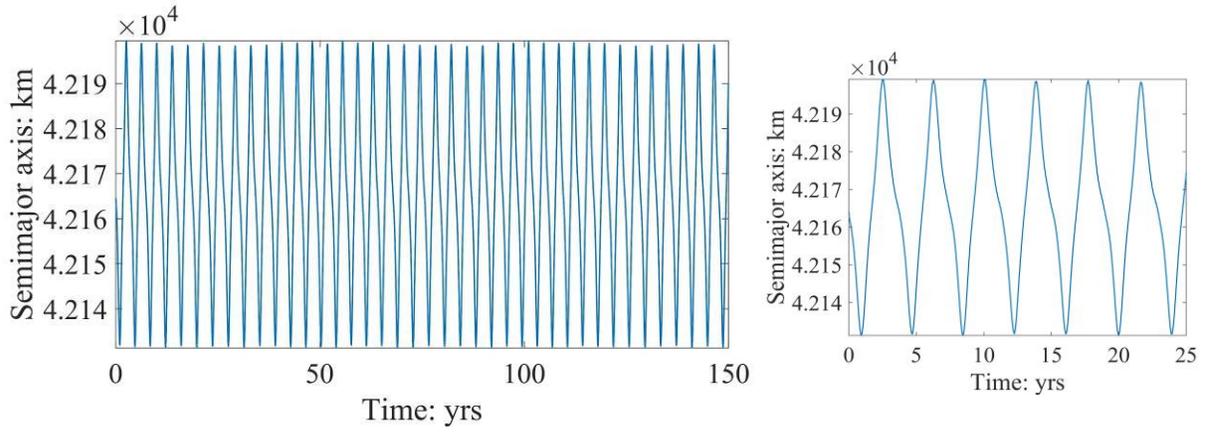

Figure 9: Time history of the semimajor axis. Detail of the first 25 years on the right.

Initial epoch: June 1st 2030, initial longitude: 0º.

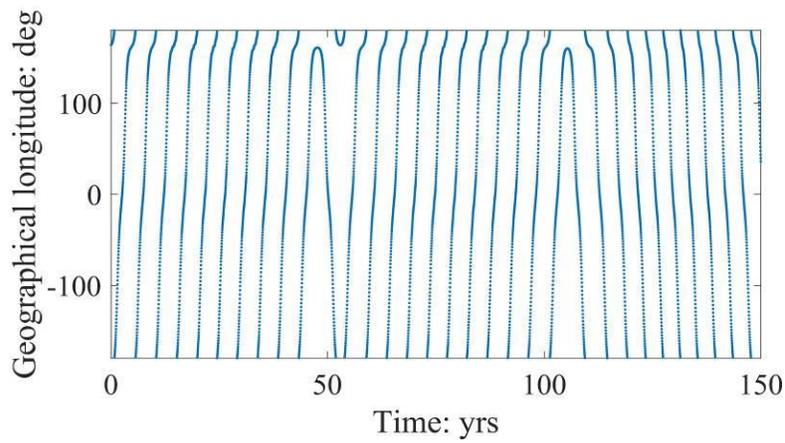

Figure 10: Time history of the geographical longitude.

Initial epoch: Jan 1st 2020, initial longitude: 164º.

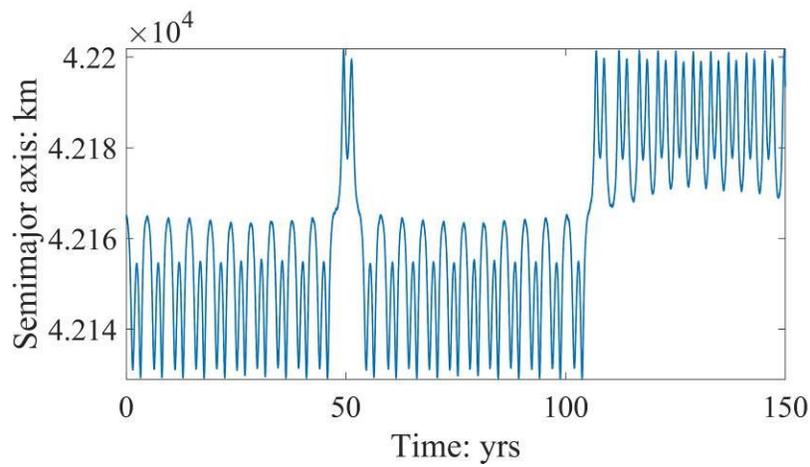

Figure 11: Time history of the semimajor axis.

Initial epoch: January 1st 2020, initial longitude: 164º.

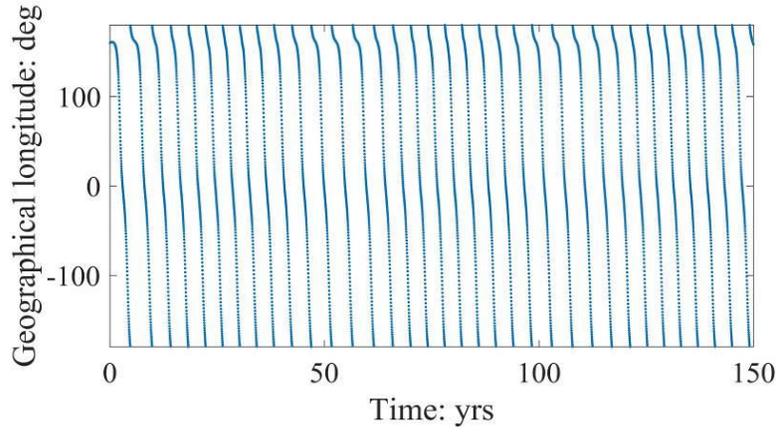

Figure 12: Time history of the geographical longitude.

Initial epoch: January 1st 2020, initial longitude: 160º.

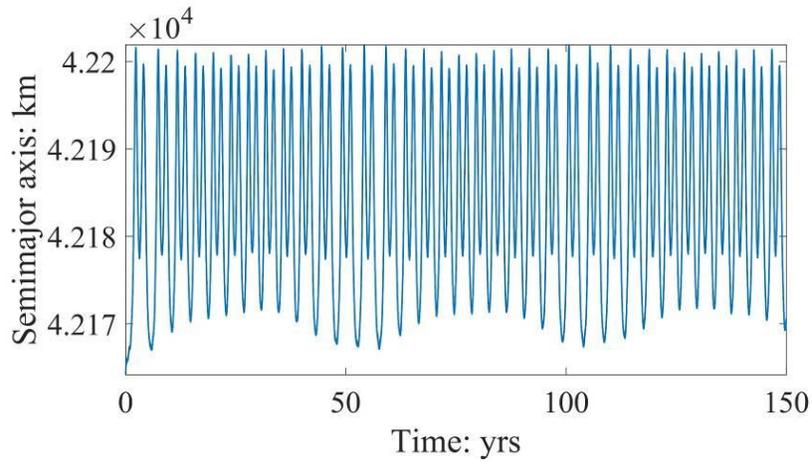

Figure 13: Time history of the semimajor axis.

Initial epoch: January 1st 2020, initial longitude: 160º.

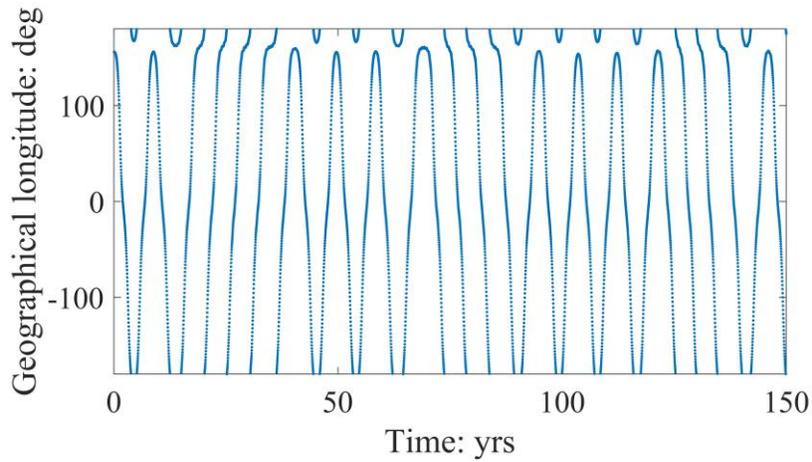

Figure 14: Time history of the geographical longitude.

Initial epoch: January 1st 2020, initial longitude: 156º, equinoctial elements propagator.

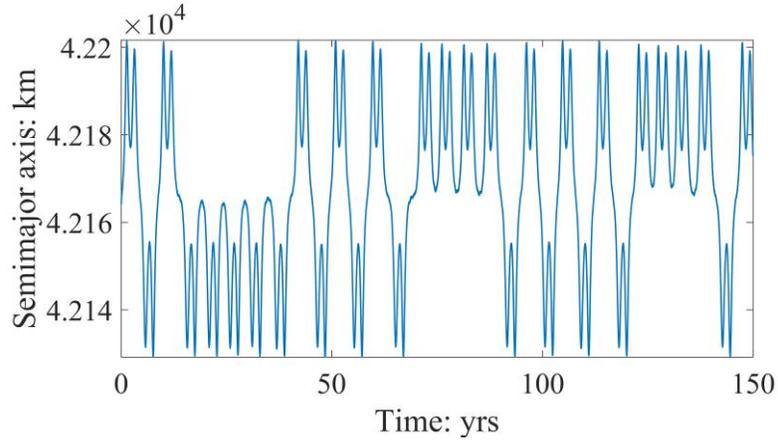

Figure 15: Time history of the semimajor axis.

Initial epoch: January 1$^{st}$ 2020, initial longitude: 156º, equinoctial elements propagator.

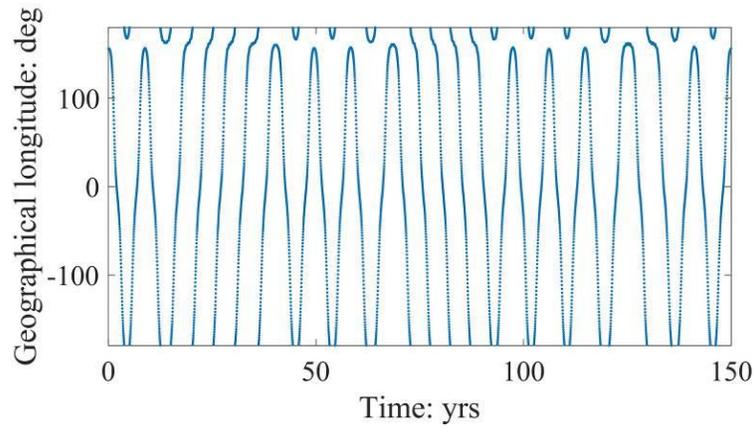

Figure 16: Time history of the geographical longitude.

Initial epoch: January 1$^{st}$ 2020, initial longitude: 156º, Cartesian propagator.

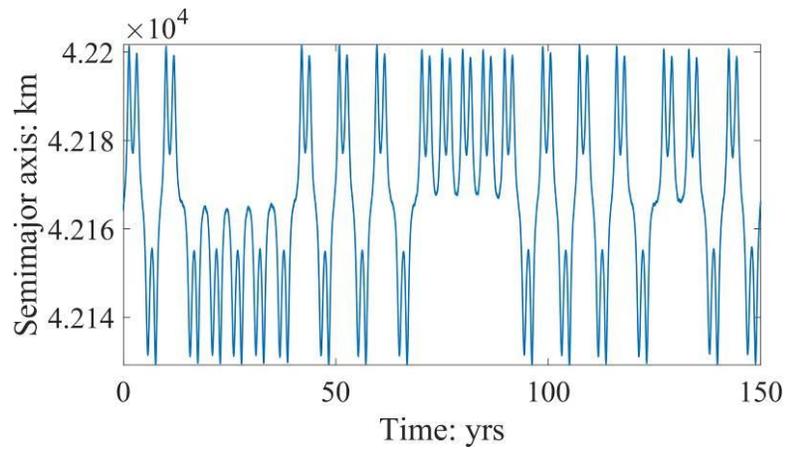

Figure 17: Time history of the semimajor axis.

Initial epoch: January 1$^{st}$ 2020, initial longitude: 156º, Cartesian propagator.

| $\lambda_g(t_0)$ | −30° | −28° | −26° | −24° | −22° | −20° | −18° | −16° |
|---|---|---|---|---|---|---|---|---|
| Behavior (R/C) | R | R | R | R | R | R | R | R |
| Type (T) | 1 | 1 | 1 | 1 | 1 | 1 | 1 | 1 |
| $\lambda_g^{(min)}$ | −173.9° | −175.3° | −176.3° | −177.7° | −178.7° | −179.5° | −180.4° | −181.0° |
| $\lambda_g^{(max)}$ | −28.6° | −26.6° | −24.6° | −22.5° | −20.5° | −18.3° | −15.6° | −13.0° |

| $\lambda_g(t_0)$ | −14° | −12° | −10° | −8° | −6° | −4° | −2° | 0° |
|---|---|---|---|---|---|---|---|---|
| Behavior (R/C) | C | C | C | C | R | R | R | R |
| Type (T) | 1-3-2 | 1-3-2 | 1-3-2 | 1-3-2 | 2 | 2 | 2 | 2 |
| $\lambda_g^{(min)}$ | −180.9° | −181.6° | −181.3° | −181.2° | −9.6° | −7.3° | −4.5° | −2.2° |
| $\lambda_g^{(max)}$ | 145.3° | 145.9° | −145.6° | 145.7° | 145.0° | 144.7° | 143.7° | 142.7° |

| $\lambda_g(t_0)$ | 150° | 152° | 154° | 156° | 158° | 160° | 162° | 164° |
|---|---|---|---|---|---|---|---|---|
| Behavior (R/C) | R | R | C | C | C | C | C | C |
| Type (T) | 3 | 3 | 3-4 | 3-4 | 3-4 | 3-4 | 3-4 | 3-4 |
| $\lambda_g^{(min)}$ | −189.7° | −191.8° | - | - | - | - | - | - |
| $\lambda_g^{(max)}$ | 153.4° | 155.7° | - | - | - | - | - | - |

| $\lambda_g(t_0)$ | 166° | 168° | 170° | 172° | 174° | 176° | 178° | 180° |
|---|---|---|---|---|---|---|---|---|
| Behavior (R/C) | C | C | R | R | R | R | R | R |
| Type (T) | 3-4 | 3-4 | 3 | 3 | 3 | 3 | 3 | 3 |
| $\lambda_g^{(min)}$ | - | - | −196.1° | −192.8° | −189.6° | −187.6° | −185.1° | −182.8° |
| $\lambda_g^{(max)}$ | - | - | 159.9° | 156.4° | 153.7° | 151.4° | 149.0° | 147.1° |

Table 1: Summary of results. Initial epoch: January 1st 2020.

| $\lambda_g(t_0)$ | −30° | −28° | −26° | −24° | −22° | −20° | −18° | −16° |
|---|---|---|---|---|---|---|---|---|
| Behavior (R/C) | R | R | R | R | R | R | R | C |
| Type ($T$) | 1 | 1 | 1 | 1 | 1 | 1 | 1 | 1-3-2 |
| $\lambda_g^{(min)}$ | −174.5 | −175.9 | −177.1 | −178.3 | −179.4 | −180.3 | −181.1 | −180.0 |
| $\lambda_g^{(max)}$ | −27.9 | −25.9 | −23.4 | −21.4 | −19.3 | −16.5 | −13.5 | 145.7 |
| $\lambda_g(t_0)$ | −14° | −12° | −10° | −8° | −6° | −4° | −2° | 0° |
| Behavior (R/C) | R | R | R | R | C | R | R | R |
| Type ($T$) | 3 | 3 | 3 | 3 | 1-3-2 | 2 | 2 | 2 |
| $\lambda_g^{(min)}$ | −180.9° | −181.6° | −181.4° | −181.2° | −9.6° | −7.3° | −4.5° | −2.2° |
| $\lambda_g^{(max)}$ | 145.3° | 145.9° | 145.5° | 145.7° | 145.0° | 144.7° | 143.7° | 142.7° |
| $\lambda_g(t_0)$ | 150° | 152° | 154° | 156° | 158° | 160° | 162° | 164° |
| Behavior (R/C) | R | R | C | C | C | R* | R* | R* |
| Type ($T$) | 3 | 3 | 3-4 | 3-4 | 3-4 | 4* | 4* | 4* |
| $\lambda_g^{(min)}$ | −190.9° | −194.7° | - | - | - | - | - | - |
| $\lambda_g^{(max)}$ | 154.7° | 158.4° | - | - | - | - | - | - |
| $\lambda_g(t_0)$ | 166° | 168° | 170° | 172° | 174° | 176° | 178° | 180° |
| Behavior (R/C) | C | C | C | R | R | R | R | R |
| Type ($T$) | 3-4 | 3-4 | 3-4 | 3 | 3 | 3 | 3 | 3 |
| $\lambda_g^{(min)}$ | - | - | - | −194.5° | −190.9° | −187.9° | −185.3° | −183.2° |
| $\lambda_g^{(max)}$ | - | - | - | 158.2° | 154.7° | 151.6° | 149.8° | 147.6° |

Table 2: Summary of the results. Initial epoch: June 1st 2030.

R* 4* indicates quasi-continuous $T$4 motion with infrequent transitions to $T$3 mode.

## 5.4 Fourier analysis

Spectral analysis offers a way to gain further insight into the mechanism behind the complex dynamics of the satellite. Initial conditions resulting in regular motion were chosen – June 1st 2030 and initial longitude 346° – as it makes the results of the Fourier analysis more representative. Figure 18 plots the semimajor axis over one complete cycle of the motion. Even though the motion is not strictly periodic due to the continuous change of the orbits of the Sun, Moon and Earth, it remains very regular over long time scales (millennia).

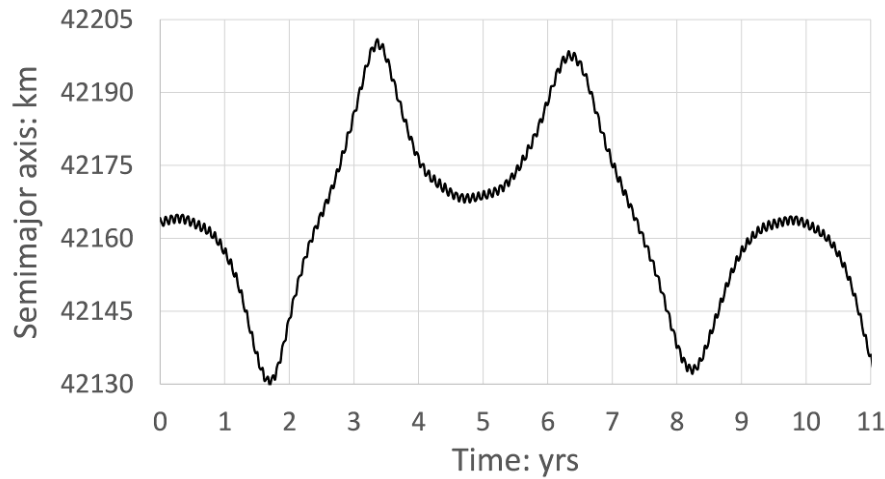

Figure 18: Time evolution of the semimajor axis over one cycle.
Initial epoch: June 1st 2030, initial longitude: 346º.

In Fig. 18 it is apparent that the fundamental component of motion has a period of 9 years. The second major harmonic has a frequency three times larger — in one 9-year cycle there are three pairs of local maxima and minima — and there is a small-amplitude modulation with a frequency of 12 year$^{-1}$, which matches the orbital period of the Moon. The reader must keep in mind that the frequencies are not general, they are functions of the initial conditions. Given that the system is nonlinear, the period of oscillation depends on the amplitude of motion. Changing the initial longitude or epoch would yield different characteristic frequencies.

Semimajor axis data sampled at 10-day intervals was passed through a 4096-point FFT (i.e., the sample length is 112 years). This yields a frequency span of 18.3 year$^{-1}$ – 1/20 day$^{-1}$ – which is sufficient to capture the fundamental harmonic of the lunar excitation. The corresponding frequency resolution is $8.92 \cdot 10^{-3}$ year$^{-1}$. The normalized complex amplitude of the Fourier transform is shown in Fig. 19.

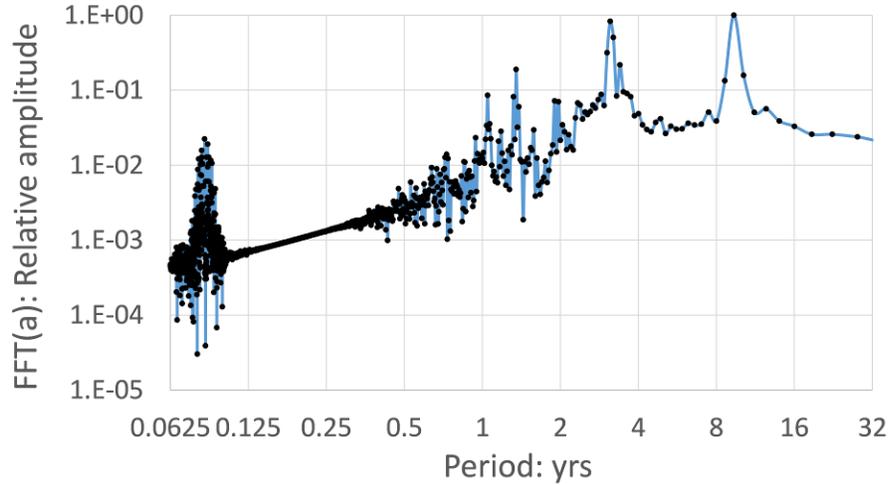

Figure 19: Fourier transform of the semimajor axis.
Initial epoch: June 1st 2030, initial longitude: 346°.

The spectrum is very noisy, which is consistent with the fact that sensitive cases are very difficult to predict over long time spans. As expected, the dominant component has a period of 9.4±0.4 years and there is a very strong harmonic at 3.15±0.05 years (3:1 ratio). Interestingly, there is an additional peak at the same 3:1 period ratio (1.05±0.005 years), albeit an order of magnitude weaker. Another standout feature is the cluster of weaker peaks centered around 1.06 months, which corresponds to the lunar excitation. The very diffuse shape of this cluster can be attributed to the highly perturbed nature of Moon's orbit, which causes a strong modulation of the excitation. The amplitude of the lunar harmonics, as expected, is comparatively small. Therefore, they become important only close to the equilibrium points of the terrestrial gravity field. Similarly, solar third body effects, typically half as large as lunar perturbations, also play an important part near the equilibrium points. Radiation pressure effects, even weaker, have a very minor role (for a spacecraft with the characteristics outlined in Section 5).

To demonstrate the dominant role of Earth's gravity, a case with the same initial conditions but removing third body effects and radiation pressure was computed and analyzed (cf. Fig. 20). The graph only displays periods above 0.25 years because, due to the absence of lunar perturbations, there is no relevant information in the lower range.

The level of noise in Fig. 20 is much lower than in Fig. 19, but the dominant peaks of the spectrogram remain virtually unchanged. This illustrates the dominance of Earth's gravitational force.

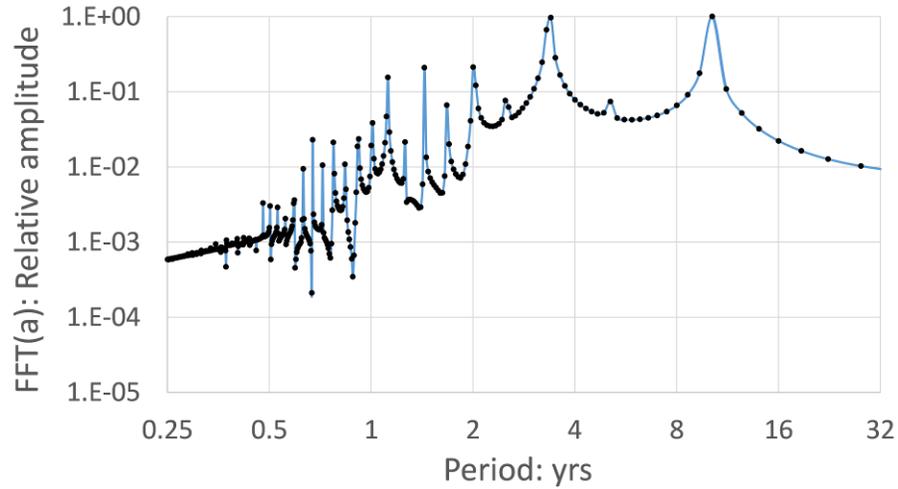

Figure 20: Fourier transform of the semimajor axis.
Initial epoch: June 1st 2030, initial longitude: 346°, Earth gravity only.

Furthermore, to demonstrate that the dynamics of the system are controlled mainly by the spherical harmonics of order below 4 – as discussed in the previous section – the analysis has been repeated using a field expansion degree of 3. The results are shown in Fig. 21. Again, the major features of the spectrogram remain unchanged, confirming our hypothesis.

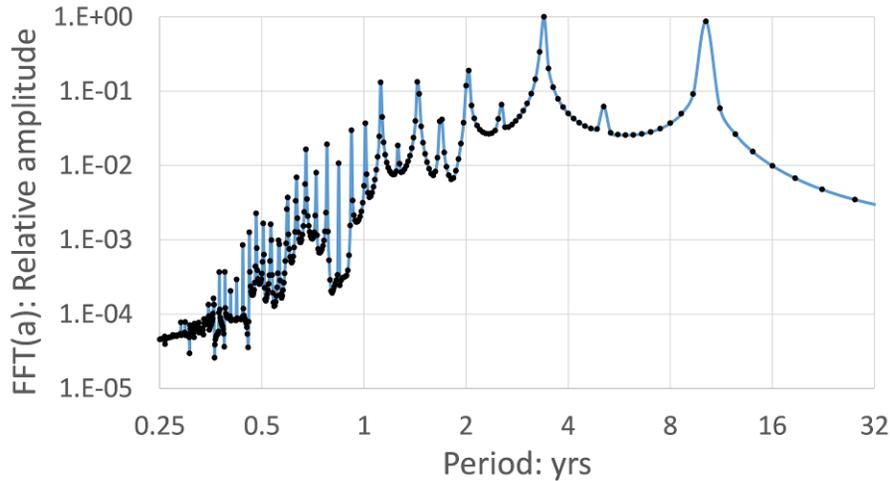

Figure 21: Fourier transform of the semimajor axis.
Initial epoch: June 1st 2030, initial longitude: 346°, Earth gravity only, expansion degree 3.

## 6. Concluding Remarks

This work revisits the long-term orbital evolution of uncontrolled GS satellites. High-fidelity numerical propagations include all the important orbital perturbations, i.e. (a) all the relevant harmonics of the geopotential, (b) gravitational pull of Sun and Moon, and (c) solar radiation pressure, which plays a minor role. Simplified analytical models (based on averaging) for the combined action of third-body perturbations and Earth oblateness predict a precession of the orbital plane about an axis tilted 7° from the pole, with a period of 52 years. We show that this motion translates into analytical relations that yield accurate predictions, i.e. cyclical changes of inclination between $0°$ and $14°$ with a corresponding variation in RAAN between $-90°$ and $+90°$.

The longitudinal dynamics has been analyzed with two different sets of equations: (i) Gauss planetary equations for modified equinoctial elements; (ii) equations of motion in Cartesian coordinates. The results from the two propagators exhibit excellent agreement over 150 years or more for most initial conditions − combinations of geographic longitude and epoch − demonstrating the reliability of the results. Only 3 scenarios out of 64 display such an extreme sensitivity that the concordance extends for less than one century. Nevertheless, the qualitative behavior remains predictable in all cases.

The longitudinal motion patterns are determined by the initial conditions. A satellite that starts close to a stable equilibrium point oscillates around it (dynamical states $T1$ and $T2$), as predicted by $J_{22}$ analytical theory. The third-degree harmonics of the geopotential create an asymmetry on the energy of the two unstable points. This gives rise to periodic motion over a wide arc ($T3$) with a small exclusion zone around the unstable point with highest energy. The time-dependent third-body gravitational effects introduce a complex modulation of the total potential that allows sporadic motion across unstable equilibrium points. Irregular transitions between modes $T1$/$T2$ and $T3$ become possible. The modulation is also responsible for migrations between $T3$ and a continuous circulation state spanning all geographical longitudes ($T4$). In some extreme cases, the satellite can be locked in a quasi-permanent East/West drift ($T4$) with very infrequent excursions to $T3$. Transitions between $T1$/$T2$ and $T4$ were not observed, due to the substantial difference in energy between the two unstable equilibrium points.

Due to the direct relationship between semimajor axis and orbital period, the radial and longitudinal dynamics are strongly correlated and follow the same complex patterns. In all the cases tested, the semimajor axis remained within ±37 km of the geostationary radius.

Spectral analysis of the satellite semimajor axis evolution (directly related to the longitudinal motion) confirmed that the dominant effects are the Earth's gravity field up to third degree harmonics and lunisolar gravitational perturbations.

We demonstrated the complexity of the longitudinal behavior of decommissioned geostationary satellites unable to perform disposal maneuvers. As a result, it deserves a thorough analysis to predict the long-term evolution of their orbits and assess the impact risk at specific longitudes.

## Acknowledgments

The work of R. Flores and E. Fantino has been supported by Khalifa University of Science and Technology's internal grants FSU-2018-07 and CIRA-2018-85. R. Flores also acknowledges financial support from the Spanish Ministry of Economy and Competitiveness, through the "Severo Ochoa Programme for Centres of Excellence in R&D" (CEX2018-000797-S).

## Appendix. Reference frames and planetary ephemerides

For the purpose of investigating long-term orbit evolution of Earth orbiting satellites, the precession motion of both the equator and the ecliptic must be included in the dynamical modeling. To do this, some useful reference frames, in conjunction with a reference epoch, are defined.

The Geocentric Celestial Reference System at the reference epoch (GCRS2000) has its origin at the Earth's center and has axes aligned with the right-hand sequence of unit vectors $(\hat{a}_1, \hat{a}_2, \hat{a}_3)$, where $\hat{a}_3$ points toward the Earth rotation axis at the reference epoch, while $\hat{a}_1$ is aligned with the intersection of the Earth equatorial plane with the ecliptic plane at the reference epoch; the latter is set to J2000.0 [38]. Due to precession motion, both the Earth equatorial plane and the ecliptic plane change. Therefore, at the initial time $t_0$ (either $t_0$ = 1 January 2020 at 0:00 UTC or $t_0$ = 1 June 2030 at 0:00 UTC in this research), these two planes identify a different reference system, named ECI($t_0$) and associated with unit vectors $(\hat{c}_1, \hat{c}_2, \hat{c}_3)$. These correspond to directions defined exactly as those of $(\hat{a}_1, \hat{a}_2, \hat{a}_3)$, with the only difference that $(\hat{c}_1, \hat{c}_2, \hat{c}_3)$ refer to the Earth equatorial plane and the ecliptic plane at $t_0$. ECI$(t_0)$ is obtained from GCRS2000 through a sequence of three elementary rotations (cf. also Fig. 22) [38],

$$[\hat{c}_1 \quad \hat{c}_2 \quad \hat{c}_3]^T = \mathbf{R}_3(-z_A(t_0))\mathbf{R}_2(\theta_A(t_0))\mathbf{R}_3(-\zeta_A(t_0))[\hat{a}_1 \quad \hat{a}_2 \quad \hat{a}_3]^T \tag{48}$$

The three angles $z_A$, $\zeta_A$, and $\theta_A$ are termed precession angles and can be evaluated using the polynomial functions of time provided by Lieske et al. [38]. The preceding relations model the long-term changes of the Earth rotation axis (precession), while neglecting short-term variations (nutation).

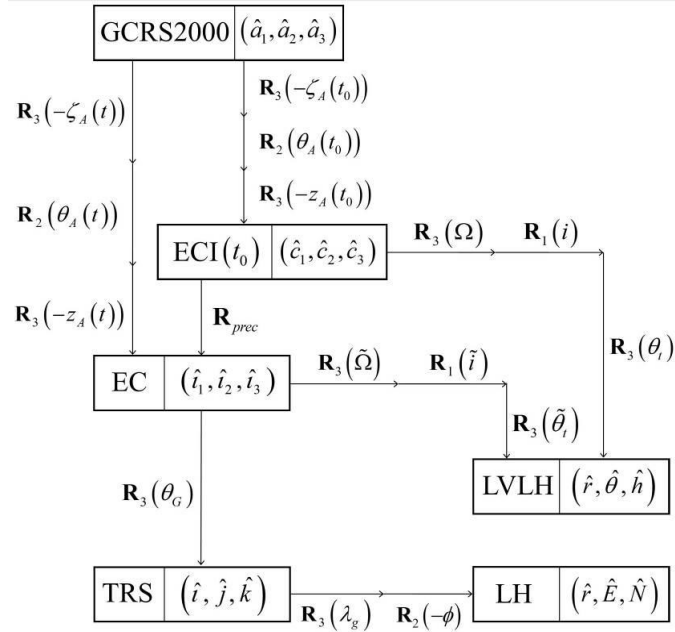

Figure 22: Graph of the relations among different reference frames.

The Earth-centered frame (EC), associated with $(\hat{i}_1, \hat{i}_2, \hat{i}_3)$, is a time-varying system that considers the instantaneous ecliptic plane and Earth equatorial plane. It is obtained from $(\hat{a}_1, \hat{a}_2, \hat{a}_3)$ through the same sequence employed for ECI$(t_0)$, with $\{\zeta_A(t), \theta_A(t), z_A(t)\}$ in place of $\{\zeta_A(t_0), \theta_A(t_0), z_A(t_0)\}$ (cf. Fig. 22),

$$[\hat{i}_1 \quad \hat{i}_2 \quad \hat{i}_3]^T = \mathbf{R}_3(-z_A(t))\mathbf{R}_2(\theta_A(t))\mathbf{R}_3(-\zeta_A(t))[\hat{a}_1 \quad \hat{a}_2 \quad \hat{a}_3]^T \tag{49}$$

Combination of Eqs. (48)-(50) leads to

$$[\hat{i}_1 \quad \hat{i}_2 \quad \hat{i}_3]^T = \underbrace{\mathbf{R}_3(-z_A(t))\mathbf{R}_2(\theta_A(t))\mathbf{R}_3(-\zeta_A(t)+\zeta_A(t_0))\mathbf{R}_2(-\theta_A(t_0))\mathbf{R}_3(z_A(t_0))}_{\mathbf{R}_{prec}}[\hat{c}_1 \quad \hat{c}_2 \quad \hat{c}_3]^T \tag{50}$$

The previous relation contains the definition of $\mathbf{R}_{prec}$. The remaining reference frames, introduced in section 2.1, are illustrated in Fig. 22 as well.

The LH-frame rotates together with the Earth and is particularly useful for the purpose of expressing the perturbing accelerations related to the harmonics of the geopotential (cf. Eq. (7)). The Greenwich sidereal time $\theta_G$ can be evaluated using the following relation [39]:

$$\begin{aligned}\theta_G = (&67310^s.54837708 + 86636^s.555367405 D_U + 6^s.9540371 \cdot 10^{-11} D_U^2 \\ &- 6^s.02 \cdot 10^{-24} D_U^3 - 1^s.122 \cdot 10^{-24} D_U^4 - 3^s.774 \cdot 10^{-32} D_U^5 + 0^s.008418264265 \Delta T) \bmod 86400^s\end{aligned} \quad (51)$$

where $D_U$ is the number of days associated to the current time since J2000 (expressed in UT1, Universal Time, directly related to the Earth's rotation), $\Delta T = \mathrm{TT} - \mathrm{UT1}$ (with TT denoting the terrestrial time), whereas $1^s = \pi/43200$ rad. $\Delta T$ measures the slowdown of the Earth rotation and is expressed in days. Over very long time periods, $\Delta T$ is expected to vary quadratically due to lunar tidal braking. However, the dynamics of Earth's interior create strong decadal variations of $\Delta T$ impossible to predict with current technology. Therefore, TT = UT1 was assumed for the calculations (likewise, JPL's Solar System Dynamics website also neglects the drift of UT1 with respect to TT for dates in the future).

For accurate evaluation of the Sun and Moon position relative to the Earth, the respective ephemerides are obtained from JPL's HORIZONS web interface [40], with a time step of 1 day. Then, they are interpolated using cubic splines that employ not-a-knot end conditions. A comparison over a time span of 150 years reveals that the positional error, defined as the difference between the position predicted by the HORIZONS system and the coordinates recovered from the interpolation algorithm, never exceeds 5 km (for the Moon) and 100 m (for the Sun).